\documentclass[twocolumn,showpacs,prl,superscriptaddress]{revtex4}

\usepackage{graphicx}
\usepackage{amssymb}

\begin{document}

\title{Glass-specific behavior in the damping of acousticlike vibrations}

\author{B. Ruffl\'e}
\author{G. Guimbreti\`ere}
\author{E. Courtens}
\author{R. Vacher}
\affiliation{Groupe de Physique des Verres et Spectroscopies, LCVN,\\
UMR 5587 CNRS, Universit\'e Montpellier II, F-34095 Montpellier Cedex 5, France}
\author{G. Monaco}
\affiliation{European Synchrotron Radiation Facility, Bo\^{\i}te Postale 220, F-38043 Grenoble, France}

\date{\today}

\begin{abstract}
High frequency sound is observed in lithium diborate glass, Li$_2$O--2B$_2$O$_3$, using 
Brillouin scattering of light and x-rays.
The sound attenuation exhibits a non-trivial dependence on the wavevector, with
a remarkably rapid increase towards a Ioffe-Regel crossover as the frequency
approaches the boson peak from below.
An analysis of literature results reveals the near coincidence of the boson-peak
frequency with a Ioffe-Regel limit for sound in {\em all} sufficiently strong glasses.
We conjecture that this behavior, specific to glassy materials, must be quite universal among them.
\end{abstract}

\pacs{63.50.+x, 78.35.+c, 81.05.Kf}

\maketitle

Phenomena occuring in glasses near the upper end of acoustic branches are of considerable interest.
Indeed, they directly relate to the thermal anomalies of these materials \cite{Zel71} and, ultimately,
to their structure on the nanometer scale.
Acoustic modes propagate in glasses as in isotropic continua up to frequencies
$\Omega /2 \pi$ of several hundred GHz \cite{Rot83}.
Their damping mainly has two origins: (i) tunneling or thermally activated relaxation (TAR) 
of defect sites \cite{Hun76}, and (ii) mode anharmonicity \cite{Vac81}.
As $\Omega /2\pi$ approaches the THz, several additional inhomogeneous-like damping mechanisms have been invoked,
such as Rayleigh scattering \cite{Gra86},
crystal-like cluster effects \cite{Duv98,Mat04}, 
or resonance with inhomogeneously distributed low-lying opticlike modes \cite{Ell01}.
We already reported extensive x-ray Brillouin-scattering investigations performed near the end of the 
longitudinal acoustic branch on permanently densified silica glass, $d$-SiO$_2$ \cite{Rat99,For02,Ruf03}.
Particular attention was given to the Brillouin linewidth $\Gamma$ obtained from
damped harmonic oscillator (DHO) fits of the spectra.
A rapid increase in $\Gamma$ is observed on approaching from below a Ioffe-Regel (IR) crossover at
$\Omega _{\rm co}$ \cite{Ruf03}.
Beyond $\Omega _{\rm co}$, the modes cease to possess a well defined wave vector $q$, strictly
implying the end of the ``branch'' $\Omega(q)$.
We also found that the acoustic excitations merge then with the boson peak (BP) \cite{For02}.
The latter corresponds to an excess in the density of vibrational states $g(\omega)$.
The position of the peak, $\Omega _{\rm BP}$, is defined by the maximum in the reduced density,
$g(\omega)/\omega ^2$.
In silica, the excess has a strong opticlike component \cite{Buc86,Tar99,Heh00}.
The relation $\Omega _{\rm co} \simeq \Omega _{\rm BP}$ implies a resonance of
BP-modes with acoustic modes near the IR-limit.
The hybridization of rather local \cite{Sur98} low lying optic modes with acoustic modes is able to
produce the IR-crossover \cite{Par01,Gur03}.
It is of course crucial to examine to what extent this property is universal in glasses.
This is the main purpose of the present Letter.

First, we report with all necessary details the rapid increase of $\Gamma$ upon approaching the IR-crossover in a
second glass, lithium diborate Li$_2$O--2B$_2$O$_3$, or LB2.
It is already known that this crossover exists in LB2 \cite{Mat01}.
It is also an excellent glass for this experiment.
Indeed, we show below that $\Omega _{\rm co}/2\pi$ is $\simeq 2.1$ THz (or $\hbar \Omega _{\rm co} \simeq 8.8$ meV)
and that the corresponding crossover wavevector $q_{\rm co}$ is $\simeq 2.1$ nm$^{-1}$.
Hence, both are sufficiently high to allow for a meaningful exploration of the approach of $\Omega _{\rm co}$
using current IXS capabilities.
The x-ray results are complemented with Brillouin light scattering (BLS) which shows that
in the GHz range, $\Gamma$ in LB2 is entirely controlled by TAR.
This is contrary to $d$-SiO$_2$, where similar measurements reveal that $\Gamma$ is dominated 
by anharmonicity \cite{Rat05}.
We find however that the onset of the IR-crossover, as well as the behavior in the crossover region,
are entirely similar in LB2 and $d$-SiO$_2$.
This indicates that the crossover mechanism is unrelated to the homogeneous broadening at lower frequencies.
Further, our results show that the BP
of LB2 also corresponds to an excess of modes, and that $\Omega _{\rm co} \simeq \Omega _{\rm BP}$.
As demonstrated below for both LB2 and $d$-SiO$_2$, plots of $\Gamma/\Omega$ {\em vs.} $\Omega$
allow to reliably obtain $\Omega _{\rm co}$.
This encouraged us, secondly, to extract from the literature on other glasses
the information on $\Gamma$ in the THz range.
Although the data are often fragmentary, plots of $\Gamma/\Omega$ 
allow in many cases to obtain meaningful estimates for $\Omega _{\rm co}$.
We find that in {\em all} instances where the BP is sufficiently strong one has $\Omega _{\rm co} \simeq \Omega _{\rm BP}$.
This is true for network glasses, for polymers, as well as for H-bonded molecular glasses.
Only for very fragile glasses in the sense of Angell \cite{Ang91},
which have quite weak BPs \cite{Sok93}, this relation might fail.

X-ray Brillouin scattering experiments on LB2 were performed on the
high-resolution spectrometer ID16 at the European
Synchrotron Radiation Facility in Grenoble, France.
The experimental conditions were similar to those described in \cite{Ruf03}.
Measurements were taken at two temperatures $T$, 300 and 573 K.
Elevated temperatures are favorable to increase the Brillouin signal by the Bose factor.
However, we refrained from nearing the glass transition, as
Li$^+$ diffusion grows then rapidly \cite{Mat02}.
Fig. 1 illustrates typical spectra at 573 K and their fits.
The indicated values of the scattering vector $Q$ correspond to the center of the collection slit
which gives a spread $\Delta Q \simeq \pm 0.18$ nm$^{-1}$.
The spectrum in Fig. 1(a) is taken at the smallest usable $Q$.
It consists of an elastic peak plus a Brillouin doublet.
The small constant background, fixed to the known detector noise, is already subtracted.
Each such spectrum was adjusted to an elastic line plus a DHO, convoluted with the separately
measured instrumental response, taking into account the frequency spread produced
by the collection aperture.
For clarity, the fitted elastic peak has been subtracted from both the data and the solid line in
Figs. 1(b-f), maintaining the original error bars.
This just makes the inelastic parts well apparent.
The Brillouin width seen in Fig. 1(b) is nearly all instrumental, while it contains a real
broadening in Fig. 1(c).
The latter becomes the dominant part in Fig. 1(d) for which $Q \sim q_{\rm co}$,
with $\hbar \Omega_{\rm co} \sim 9$ meV as explained below.
It is seen that the DHO lineshape starts deviating systematically from the measured
signal around $q_{\rm co}$.
The spectrum in Fig. 1(e) is actually the mean of four spectra accumulated
from 4.1 to 4.4 nm$^{-1}$.
At these high $Q$-values the spectra evolve so slowly with $Q$ that taking a mean is a good
procedure for increasing the signal-to-noise ratio.
The additional spread in $Q$ was taken into account in the fit which however is then quite poor, showing that the DHO
is no more a valid approximation.
All these observations are remarquably similar to previous ones on $d$-SiO$_2$.
Fig. 1(f) shows a spectrum at 29.4 nm$^{-1}$ fitted to a log-normal \cite{Mal90}. 
This is essentially the BP as discussed below.
\begin{figure}
\includegraphics[width=8.5cm]{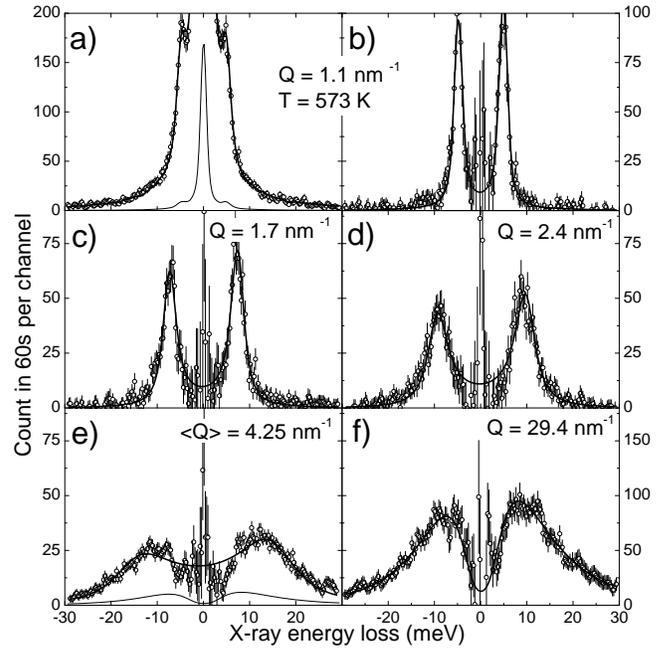}
\caption{IXS spectra of LB2 and their fits explained in the text;
(a) A full spectrum and its central portion $\times 1/20$;
(b-e) The inelastic part remaining after subtraction of the elastic component determined in a global fit
of each spectrum.
The lines show the DHO adjustment in this global fit;
(f) Inelastic part of a high-$Q$ spectrum treated in the same manner but
adjusted to a log-normal reproducing the BP.}
\end{figure}

\begin{figure}
\includegraphics[width=8.5cm]{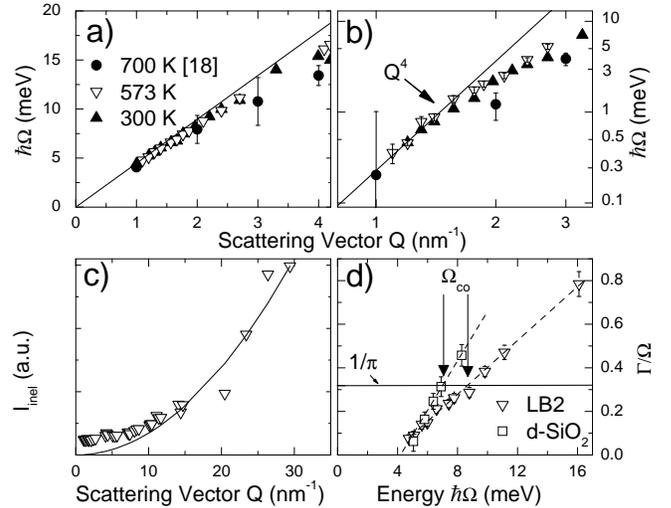}
\caption{(a-b) The parameters of the DHO fits for the two temperatures 573 and 300 K. The three full dots
at 700 K are from \cite{Mat01}. The slope of the line in (a) is the BLS velocity measured independently. In
(b), the error bars at 573 K are mostly smaller than the symbols, while these at 300 K (not shown) are about
twice as large. The line of slope 4 in (b) is drawn through the first five points at 573 K. (c) The integrated 
inelastic intensities at 573 K, corrected for the oxygen-ion form factor, with a line $\propto Q^2$.
(d) The method used to determine $\Omega _{\rm co}$ applied to LB2 (these data) and $d$-SiO$_2$ (data from
\cite{Ruf03}). The straight lines are guides to the eye. The crossover positions are at the intercept with
the ordinate $1/\pi$.}
\end{figure}
The parameters extracted from the DHO fits of the IXS spectra are shown in Fig. 2.
They exhibit very little dependence on $T$, if any.
Below the crossover there is only a slight departure from linear dispersion as seen in Fig. 2(a).
In this region one can consider that the experiment detects approximate plane waves with $q=Q$.
There is however a very rapid increase of the full width $\Gamma$ at the lowest $Q$ values, as
shown in Fig. 2(b).
A power law $Q^4$ has been drawn through the five lowest points at 573 K.
A fit to these five points actually gives the power $4.2 \pm 0.1$.
As can be expected, this rapid increase of $\Gamma$ saturates as the crossover
is approached from below, starting here for $Q \sim 2$ nm$^{-1}$.
There is then a progressive merging of the sound excitations with the opticlike BP.
Although $Q$ remains fixed by the scattering geometry, the excitations that are
observed might then be so far from plane waves that $q$ becomes ill-defined.
The position of opticlike modes is $Q$-independent, which we have observed in comparing
spectra taken at 23.4, 26.4, and 29.4 nm$^{-1}$.
In the incoherent approximation, 
their integrated intensity should be proportional to $Q^2$ \cite{Buc85}.
The total inelastic intensity $I_{\rm inel}$, corrected for the atomic form factor of oxygen,
is shown in Fig. 2(c).
At low $Q$, the Brillouin signal contributes a constant.
The $Q^2$ dependence is seen at high $Q$, confirming that Fig. 1(f) shows mainly the scattering from the BP,
whose shape actually agrees with Raman scattering \cite{Lor84}.
Extrapolating the intensity of the log-normal shown in Fig. 1(f) to the $Q$-value of Fig. 1(e) leads to
the thin line drawn there.
This is only a small fraction of the spectral intensity, so that
what is seen in Figs. 1(b-e) is mostly produced by acousticlike excitations while at higher $Q$
these merge into opticlike modes.
A similar merging of the acoustic mode into the BP was observed in $d$-SiO$_2$ \cite{For02}.
Incidentally, we also show in Fig. 2(a-b) the three data points from the previous
measurement \cite{Mat01}.
Besides a slight quantitative difference, it is obvious that there is
there insufficient information to allow for the present discussion.

It remains to determine $\Omega_{\rm co}$.
The precise location of the IR-limit is a matter of definition.
We place the crossover at the frequency where the energy mean free path of sound
first reaches down to half its wavelength, $\ell = \lambda/2$.
The simple exponential decay model leading to the DHO gives $\ell ^{-1} = \Gamma / v$, where
$\Gamma$ is in rad/s and $v=\Omega /q$ is the sound velocity.
With $q=2\pi/\lambda$, these relations lead to the crossover criterion $\Gamma=\Omega/\pi$.
Although a peak in the DHO spectrum is still clearly identified for $\Gamma=\Omega/\pi$, one 
recognizes that a plane wave whose energy decays by a factor $1/e^2$ over a wavelength corresponds well
to the concept of a IR-limit \cite{Iof60}.
Fig. 2(d) shows plots of $\Gamma/\Omega$ {\em vs.} $\Omega$ for LB2 and $d$-SiO$_2$.
The intercept with the horizontal line at $1/\pi$ gives $\Omega_{\rm co}$.
For LB2, we find $ \hbar \Omega _{\rm co} \simeq 8.8 \pm 0.2$ meV and $q _{\rm co} \simeq 2.1$ nm$^{-1}$, whereas
for $d$-SiO$_2$ one has with this criterion 
$ \hbar \Omega _{\rm co} \simeq 7.0 \pm 0.2$ meV and $q _{\rm co} \simeq 1.8$ nm$^{-1}$.
In both cases $\Omega _{\rm co} \simeq \Omega _{\rm BP}$ as illustrated in Fig. 3.

\begin{figure}
\includegraphics[width=8.5cm]{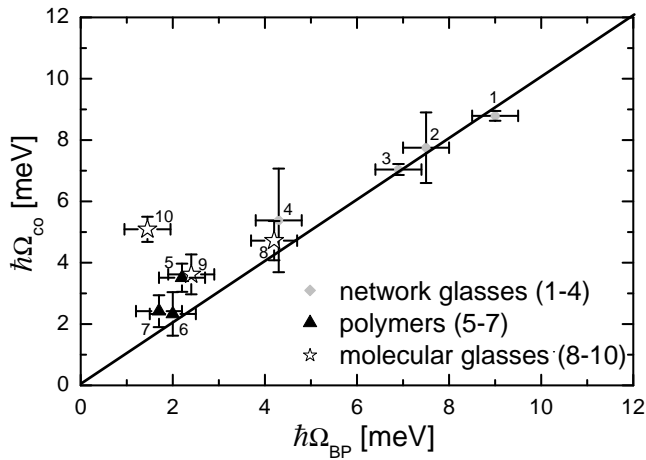}
\caption{$\Omega _{\rm co}$ {\em vs.} $\Omega_{\rm BP}$ for various glasses.
The line $\Omega _{\rm co}= \Omega_{\rm BP}$ is a guide to the eye.
The points and the references for $\Gamma (\Omega)$ and $\Omega_{\rm BP}$ are:
(1) Li$_2$O--2B$_2$O$_3$, here and \cite{Lor84,Mat01};
(2) Li$_2$O--4B$_2$O$_3$ \cite{Lor84,Mat01};
(3) $d$-SiO$_2$ \cite{Ruf03,For02};
(4) vitreous silica $v$-SiO$_2$ \cite{Del98,Heh00};
(5) polybutadiene \cite{Fio99,Buc96};
(6) polycarbonate \cite{Mat03};
(7) Se \cite{Sco04,For98};
(8) glycerol \cite{Set98,Wut95};
(9) ethanol \cite{Mat04,Ram97};
(10) OTP \cite{Mon98,Toe01}.}
\end{figure}
Although $\Gamma/\Omega \propto \Omega^3$ in the onset of the crossover, straight lines
provide excellent approximations to $\Gamma/\Omega$ {\em vs.} $\Omega$ 
over a relatively broad range around $\Omega_{\rm co}$, as seen in Fig. 2(d).
This is observed both in LB2 and $d$-SiO$_2$.\cite{foot1}
%As the straight lines do not cross the origin, this does not imply $\Gamma \propto \Omega^2$.
It is reasonable to expect that other glasses will have a similar behavior.
Even though the crossover was mostly not investigated by others, sufficient information
is available on several glasses to draw plots of $\Gamma/\Omega$.
A summary of such an analysis is presented in Fig. 3.
The vertical error bars result from the indetermination in the published $\Gamma (\Omega )$ data.
We kept in Fig.3 {\em all} glasses for which these errors are reasonably small.
For example, the information on CKN \cite{Mat01b} is unfortunately not sufficiently
accurate to be included in this plot.
The horizontal error bars, the same for all points, are a fair estimate of the inaccuracy in $\Omega_{\rm BP}$. 
Fig. 3 demonstrates that indeed $\Omega _{\rm co} \simeq \Omega _{\rm BP}$ in the vast majority of cases.
Analogous considerations were already presented in \cite{Par01}, based on
estimates of $\Omega_{\rm co}$ using the soft-potential model combined with measured
two-level-system parameters.
Here, the same conclusion is reached based on direct measurements.
A plausible explanation for this widespread behavior is that a bilinear coupling of quasilocal oscillators
with the strain field redistributes the density of states of the former, leading to
$\Omega _{\rm BP} \simeq \Omega _{\rm co}$ \cite{Par01,Gur03}.
Out of the ten glasses in Fig. 3, only OTP falls appreciably off the line.
It is known that very fragile glasses, like OTP, have very weak BPs \cite{Sok93}, in which case the
hybridization might not suffice to produce a clear IR-crossover.
This seems to be also the case for CKN, another very fragile glass, where a crossover effect seems
absent from the available data \cite{Mat01b}.

Given the universality $\Omega _{\rm co} \simeq \Omega _{\rm BP}$ in sufficiently
strong glasses, it is natural to expect that the related rapid growth of $\Gamma$
below crossover should also be universal.
If so, in all these glasses there should be a region where $\Gamma$ increases approximately with
$\Omega^4$, or also with $Q^4$ since there $Q \propto \Omega$ \cite{foot2}.
Hence, the belief that there is a ``universal'' law $\Gamma \propto Q^2$ in glasses \cite{Mat01b,Ruo99}
should be questioned, at least below $\Omega _{\rm co}$.
To illustrate this point we explored $\Gamma$ over a larger frequency range in LB2.
The results are summarized in Fig. 4 on the energy scale usually used in IXS.
BLS was excited with an Ar-ion laser at 514.5 nm and analyzed with a high resolution
tandem spectrometer that includes a spherical Fabry-Perot as described in \cite{Vac80}. 
The Brillouin linewidth was measured in function of the scattering angle and temperature.
The results are illustrated for three geometries and two $T$-values in Fig. 4.
The error bars are smaller than the symbols.
A line of slope 1 is drawn through the points at 573~K as a guide to the eye.
This linear dependence, $\Gamma \propto \Omega$, is characteristic of TAR at frequencies $\Omega$
near the relaxation maximum, that is for $\Omega \tau \sim 1$ where $\tau$ is a typical TAR time.
On the other hand, at ultrasonic (US) frequencies, the absorption is close to the regime $\Gamma \propto \Omega^2$,
as shown by the dashed line of slope 2. 
This is expected for TAR in the $\Omega \tau \ll 1$ regime.
\begin{figure}
\includegraphics[width=8.5cm]{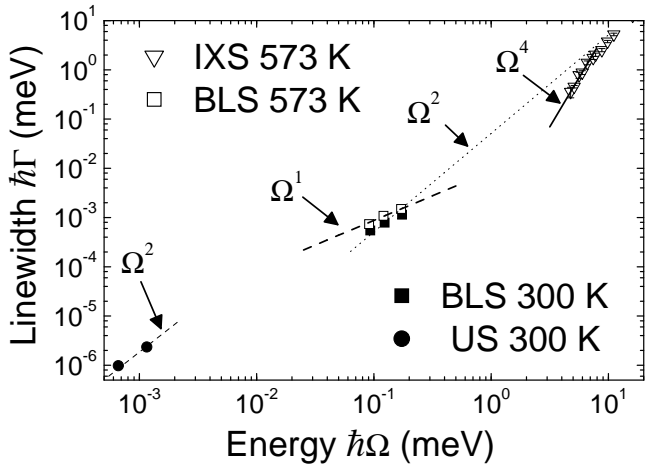}
\caption{$\Gamma (\Omega)$ for LB2, showing the IXS results with the line of slope 4 from Fig. 2(b), 
our own BLS measurements, and ultrasonic (US) values from \cite{Cip81}.
The lines are guides to the eye discussed in the text.} 
\end{figure}
Thus in LB2, the observed damping in the MHz and GHz ranges seems all controlled by TAR.
The anharmonic damping contribution, $\Gamma_{\rm anh} \propto \Omega^2$, can be estimated using the
$T$-dependence of the Brillouin sound velocity, $\delta v/v = -6.5 \times 10^{-5} \; T$(K),
as explained in \cite{Vac05}.
It is known that such a decrease of $v$ is characteristic of anharmonic
coupling to the thermal bath \cite{Cla78}.
Taking as rough estimate for the thermal-bath relaxation time that of $d$-SiO$_2$ \cite{Rat05}, one finds
that $\Gamma_{\rm anh}$ in the BLS range should be about an order of magnitude smaller than the observed $\Gamma$.
In Fig. 4, the dotted line in $\Omega ^2$  is only drawn to point out that such an interpolation between BLS and IXS
results is not justified.
The $Q^2$ behavior often reported above crossover might relate to what amounts to a strong inhomogeneous
broadening as suggested in \cite{Mat04}.
It is meaningless to extrapolate it below the crossover.   

The central result of this study, that $\Omega _{\rm co} \simeq \Omega _{\rm BP}$ in sufficiently strong glasses,
conversely implies that the observation of a sizable BP indicates the presence of a IR-crossover.
One expects then $q_{\rm co} \simeq \Omega _{\rm BP}/v$  which is always rather small.
For example, in LB2, we found here that $2\pi/q_{\rm co} \simeq 30$ \AA.
A cube of that size contains over 200 formula units, Li$_2$B$_4$O$_7$.
It is obvious that most acousticlike excitations are of shorter characteristic length, although
these might be far from plane waves.
A signature of such modes was actually observed in silica glass \cite{Ara99}.
It appears on the neutron-scattering inelastic structure factor, $S(Q,\omega)$.
The maxima in $\omega$ of $S(Q,\omega)$ form {\em pseudo}-branches, $\Omega(Q)$.
The latter are crystal-like, the peaks in the elastic structure factor $S(Q)$ playing the role
of Bragg reflections.
Hence, it is not so surprising that comparing a glass to the corresponding polycrystal,
acoustic measurements at very high $Q$ give similar pseudo-branches and
inhomogeneous linewidths, as reported for ethanol in \cite{Mat04}.
The glass-specific structural and dynamical properties should not be sought on the nearest neighbor scale, but
rather at the nanometer scale corresponding to $q_{\rm co}$.

The authors thank A. Matic for the LB2 sample, R. Vialla for recent improvements to the BLS spectrometer,
and J.M. Fromental for technical assistance.


\begin{references}
\bibitem{foot1}As the straight lines do not cross the origin, this does not imply 
$\Gamma \propto \Omega^2$.
\end{references}
\end{document}